\documentclass[a4paper,12pt]{article}
\usepackage{epsfig}
\usepackage{amssymb}
\usepackage{feynmf}
\usepackage{axodraw}

\newcommand{\ie}{{\it i.e.}}

\newcommand{\mrm}[1]{\mbox{\rm #1}}
\newcommand{\be}{\begin{equation}}
\newcommand{\ee}{\end{equation}}
\newcommand{\br}{\begin{eqnarray}}
\newcommand{\bea}{\begin{eqnarray}}
\newcommand{\eea}{\end{eqnarray}}
\newcommand{\er}{\end{eqnarray}}
\newcommand{\ba}{\begin{array}}
\newcommand{\ea}{\end{array}}
\newcommand{\bi}{\begin{itemize}}
\newcommand{\ei}{\end{itemize}}
\newcommand{\bn}{\begin{enumerate}}
\newcommand{\en}{\end{enumerate}}
\newcommand{\bc}{\begin{center}}
\newcommand{\ec}{\end{center}}

\newcommand{\eps}{\epsilon}

\newcommand{\nn}{\nonumber\\}

\newcommand{\rfn}[1]{(\ref{#1})}

\newcommand{\gsim}{\lower.7ex\hbox{$\;\stackrel{\textstyle>}{\sim}\;$}}
\newcommand{\lsim}{\lower.7ex\hbox{$\;\stackrel{\textstyle<}{\sim}\;$}}

\begin{document}
\tolerance=100000
\thispagestyle{empty}
\setcounter{page}{0}

\begin{flushright}
{\rm CERN-TH/2002-239} \\
{\tt hep-ph/0209207}
\end{flushright}

\vspace*{\fill}

\begin{center}
{\Large \bf 
Higgs-Mediated $B_{s,d}^0\to\mu\tau, e\tau$ and $\tau\to3\mu, e\mu\mu$ 
Decays in Supersymmetric Seesaw Models
}\\[2.cm]

{{\large\bf Athanasios Dedes}$^1$, 
{\large\bf John Ellis}$^2$ {\large and} {\large\bf Martti 
Raidal}$^{2,3}$}  
\\[7mm]

{\it $^1$Physikalisches Institut der Universit\"at Bonn, 
 Nu\ss allee 12,  D-53115 Bonn, Germany
} \\[3mm]

{\it $^2$ TH Division, CERN, CH-1211 Geneva 23, Switzerland
} \\[3mm]

{\it $^3$ National Institute of Chemical Physics and Biophysics, 
Tallinn 10143, Estonia
} \\[10mm]
\end{center}

\vspace*{\fill}

\begin{abstract}
{\small\noindent
We study the rates allowed for the Higgs-mediated decays
$B_{s,d}^0\to\mu\tau, e\tau$ and $\tau\to \mu\mu\mu, e\mu\mu$ in
supersymmetric seesaw models, assuming that the only source of lepton
flavour violation (LFV) is the renormalization of soft
supersymmetry-breaking terms due to off-diagonal singlet-neutrino Yukawa
interactions. These decays are strongly correlated with, and constrained
by, the branching ratios for $B_{s,d}^0\to\mu\mu$ and $\tau\to
\mu(e)\gamma.$ Parametrizing the singlet-neutrino Yukawa couplings $Y_\nu$
and masses $M_{N_i}$ in terms of low-energy neutrino data, and allowing
the flavour-universal soft masses for sleptons and for squarks, as well 
as those for the two
Higgs doublets, to be different at the unification scale, we 
scan systematically over the model
parameter space. Neutrino data and the present experimental constraints
set upper limits on  the Higgs-mediated LFV decay rates
$Br(B_{s}^0\to\mu\tau,  e\tau)\lsim 4\times 10^{-9}$ and $Br(\tau\to
\mu\mu\mu, e\mu\mu)\lsim 4\times 10^{-10}$.
}
\end{abstract}

\vspace*{\fill}

\begin{flushleft}
{\rm CERN-TH/2002-239}\\
{\rm  September 2002} \\
\end{flushleft}

\newpage
\setcounter{page}{1}


{\it 1. Introduction.} 
In the minimal supersymmetric extension of the Standard Model (MSSM),
non-holomorphic Yukawa interactions of the form $D^c Q H_2^*$ are
generated at one-loop level~\cite{cgnw}. At large values of
$\tan\beta=v_2/v_1$, the contribution of such loop-suppressed operators to
the down-type quark masses may become comparable to those from the usual
superpotential terms $D^c Q H_1.$ As these two contributions have 
different
flavour structures, because of the up-type quark Yukawa interactions, they
cannot be diagonalized simultaneously~\cite{Maxim,Babu}.
This difference leads to potentially
large new contributions to Higgs-mediated flavour-changing processes
involving down-type quarks, such as 
$B_{s,d}^0\to\mu\mu$~\cite{Bobeth,Isidori,UMSSM,SUSYbreakingmodels,Buras} and
$B^0$-$\bar B^0$ mixing~\cite{Buras,Isidori}.

This discussion cannot be generalized directly to the charged-lepton
sector, because the MSSM has no right-handed neutrinos at the electroweak
scale. However, experimental data convincingly indicate that neutrinos do
have masses~\cite{mn}. 
Their small values are most naturally explained via the
seesaw mechanism~\cite{seesaw}, 
which involves super-heavy singlet (right-handed)
neutrinos with masses $M_{N_i}$. In this case, the presence of neutrino
Yukawa couplings $N^c_i(Y_\nu)_{ij}L_jH_2$ above the heavy-neutrino
decoupling scale induces off-diagonal elements in the left-slepton mass
matrix via renormalization~\cite{bm,h1},
\begin{eqnarray}
(\Delta m_{\tilde{L}}^2)_{ij}&\simeq&
-\frac{1}{8\pi^2}(3m_0^2+A_0^2) 
(Y^\dagger L Y)_{ij} \,, ~~ L=\ln\frac{M_{GUT}}{M_{N_i}} \delta_{ij}\,, 
\label{leading}
\end{eqnarray}
even if the initial slepton soft supersymmetry-breaking masses $m_0$ and
trilinear couplings $A_0$ are flavour-universal at $M_{GUT}.$ This is the
only source of lepton-flavour violation (LFV) in supersymmetric seesaw 
models with flavour-universal
soft mass terms. Since it is induced by heavy-neutrino Yukawa
interactions, it relates the LFV processes to low-energy
neutrino data. At the one-loop level, (\ref{leading}) also gives rise to
flavour violation in non-holomorphic interactions of the form $E^c L H_2^*$
and leads to Higgs-mediated LFV processes in the charged-lepton sector.

\begin{figure}[t]
\begin{center}
\begin{picture}(320,80)(0,0)
\ArrowLine(50,40)(20,70)
\ArrowLine(20,10)(50,40)
\Vertex(50,40){5}
\ArrowLine(140,10)(110,40)
\ArrowLine(110,40)(140,70)
\DashLine(50,40)(110,40){3}
\Vertex(110,40){2}
\Text(80,30)[]{\small $h^0$,$H^0$,$A^0$}
\Text(35,70)[]{\small $b_R$}
\Text(40,10)[]{\small $s_L,d_L$}
\Text(125,70)[]{\small $\mu$}
\Text(125,10)[]{\small $\mu$}
\Text(17,42)[]{\small $\tan^2\beta$}
\Text(140,40)[]{\small $m_\mu\tan\beta$}
\Text(80,-10)[]{\small \bf (a)}
\ArrowLine(220,40)(190,70)
\ArrowLine(190,10)(220,40)
\Vertex(220,40){5}
\ArrowLine(310,10)(280,40)
\ArrowLine(280,40)(310,70)
\DashLine(220,40)(280,40){3}
\Vertex(280,40){2}
\Text(250,30)[]{\small $h^0$,$H^0$,$A^0$}
\Text(205,70)[]{\small $\tau_R$}
\Text(210,10)[]{\small $\mu_L,e_L$}
\Text(295,70)[]{\small $\mu$}
\Text(295,10)[]{\small $\mu$}
\Text(188,42)[]{\small $\tan^2\beta$}
\Text(310,40)[]{\small $m_\mu\tan\beta$}
\Text(250,-10)[]{\small \bf (b)}
\end{picture}
\end{center}
\caption{\it 
Dominant Higgs penguin diagrams contributing to (a) $B^0_{s,d}\rightarrow 
\mu\mu$ 
and (b) $\tau\to\mu\mu\mu$ decays at large $\tan\beta$.}
\label{fig1}
\end{figure}
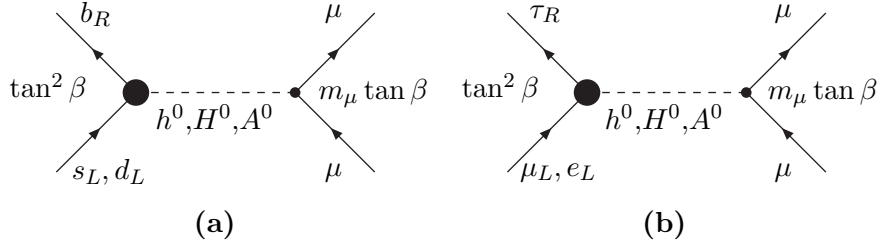

It is well known that, at large $\tan\beta$, new Higgs-mediated
contributions to the decay $B_{s,d}^0\to\mu\mu$ may exceed the Standard
Model (SM) contribution by orders of magnitude. The dominant contributions
come from the diagrams in Fig.~\ref{fig1}~(a) which are determined by
Cabibbo-Kobayashi-Maskawa (CKM) matrix elements if the squark mass
matrices are flavour-universal at the GUT scale, as suggested by the data
on flavour-changing neutral interactions~\cite{ENBG}. It was suggested
recently~\cite{Babu2} that the Higgs-mediated contribution to the LFV
process $\tau\to\mu\mu\mu$ depicted in Fig.~\ref{fig1}~(b) might be
sizeable in supersymmetric seesaw models~\footnote{
For a discussion of LFV decays in non-supersymmetric models the reader 
is referred to~\cite{Apostolos,Sher}.}, with a branching ratio as large
as $Br(\tau\to\mu\mu\mu)\sim {\cal O}(10^{-7})$. This claim was made without
a complete study of all related LFV processes.

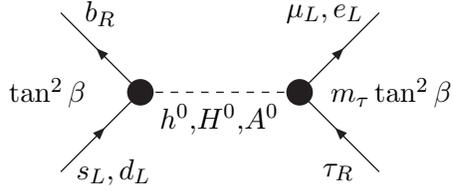
\begin{figure}[t]
\begin{center}
\begin{picture}(160,80)(0,0)
\ArrowLine(50,40)(20,70)
\ArrowLine(20,10)(50,40)
\Vertex(50,40){5}
\ArrowLine(140,10)(110,40)
\ArrowLine(110,40)(140,70)
\DashLine(50,40)(110,40){3}
\Vertex(110,40){5}
\Text(80,30)[]{\small $h^0$,$H^0$,$A^0$}
\Text(35,70)[]{\small $b_R$}
\Text(40,10)[]{\small $s_L,d_L$}
\Text(120,70)[]{\small $\mu_L,e_L$}
\Text(125,10)[]{\small $\tau_R$}
\Text(15,40)[]{\small $\tan^2\beta$}
\Text(145,40)[]{\small $m_\tau\tan^2\beta$}
\end{picture}
\end{center}
\caption{\it
Dominant Higgs penguin diagrams contributing to $B^0_{s,d}\rightarrow 
\tau\mu$ 
decays at large $\tan\beta$.}
\label{fig2}
\end{figure}

Since the only source of LFV in this model is (\ref{leading}), and all LFV
processes are induced by slepton-neutralino and sneutrino-chargino loops,
Higgs-mediated LFV is constrained indirectly by limits on the
processes $\tau\to\mu\gamma$, $\tau\to e\gamma$ and $\mu\to e\gamma,$
which have been  well studied by experiment.
 Although $\tau\to\mu(e)\gamma$ and Higgs-mediated LFV 
depend on the model parameters in systematically different ways, 
the former impose important and unavoidable constraints on the latter 
processes. Additional constraints are expected to come from
lepton-flavour-conserving but quark-flavour-violating processes such as
$B_s^0\to\mu\mu$, which depends on Higgs boson masses in the same way as
Higgs-mediated $\tau\to\mu\mu\mu$ (see Fig.~\ref{fig1}). In addition,
there are other Higgs-mediated LFV processes such as $B_s^0\to\mu\tau$ and
$B_s^0\to e\tau$, which are induced by double-penguin diagrams, as
depicted in Fig.~\ref{fig2}, and may also be of experimental interest.
The current experimental upper limits on such decays are shown in 
Table~\ref{tab:BLFV}.
Although the double-penguin diagram is suppressed by an additional loop
factor, its amplitude is enhanced by $m_\tau/m_\mu\tan\beta$, as seen by
comparing Fig.~\ref{fig2} with Fig.~\ref{fig1}. Therefore, at large 
$\tan\beta,$ the rate for
$B_s^0\to\mu\tau$ may even exceed that of $\tau\to\mu\mu\mu.$

%
%

To quantify these qualitative statements, we adopt for the moment the
approximation considered in~\cite{Babu2}, \ie, we take all the
supersymmetry-breaking mass parameters of the model to be equal at low
scales, use heavy-neutrino masses that are degenerate with $M_N=10^{14}$
GeV, and assume that $(Y_\nu^\dagger Y_\nu)_{32,33}=1.$ This approximation is
not realistic, but it is useful for comparing the sensitivities of
different processes to new physics: a more complete treatment of these
processes is presented in the following Sections of this Letter. In the
simplified case, working in the mass-insertion approximation
and following exactly~\cite{Babu2}, we obtain
\bea
Br(\tau\to3\mu)\simeq 1.6\times 10^{-8}\left[\frac{\tan\beta}{60}\right]^6
\left[\frac{100\,\mrm{GeV}}{M_A}\right]^4 \,,
\label{t3map}
\eea
which is about six times smaller than the estimate quoted in that paper.
This should be compared with the corresponding estimate 
\bea
Br(\tau\to\mu\gamma)\simeq 
1.3\times 10^{-3}\left[\frac{\tan\beta}{60}\right]^2
\left[\frac{100\,\mrm{GeV}}{M_S}\right]^4 \,.
\label{tmgap}
\eea
Both equations \rfn{t3map} and \rfn{tmgap} are valid in the
large-$\tan\beta$ limit. Whereas \rfn{t3map} is two orders of magnitude
{\it below} the present experimental bound on $\tau\to3\mu$, \rfn{tmgap} 
is three orders of magnitude {\it above} the present bound on
$\tau\to\mu\gamma.$ There is also the photonic penguin contribution to the
decay $\tau\to3\mu$, which is related to $Br(\tau\to\mu\gamma)$ by
\bea
Br(\tau\to3\mu)_{\gamma}=\frac{\alpha}{3\pi}
\left(\ln\frac{m^2_\tau}{m^2_\mu}-\frac{11}{4}\right) Br(\tau\to\mu\gamma).
\eea 
Numerically, \rfn{tmgap} leads to 
\bea
Br(\tau\to3\mu)_\gamma\simeq 
3.0\times 10^{-6}\left[\frac{\tan\beta}{60}\right]^2
\left[\frac{100\,\mrm{GeV}}{M_S}\right]^4 \,,
\label{ph}
\eea
which is a factor of 100 larger than \rfn{t3map}. Notice also that
suppressing \rfn{tmgap} by postulating large slepton masses would suppress
\rfn{t3map} at the same time, since sleptons enter into both loops.
However, suppressing \rfn{t3map} by a large Higgs mass $M_A$ would not
affect $\tau\to\mu\gamma.$

With the same assumptions, we obtain 
\bea
Br(B_s^0\to\mu\mu)\simeq 1.9\times 10^{-5}\left[\frac{\tan\beta}{60}\right]^6
\left[\frac{100\,\mrm{GeV}}{M_A}\right]^4 \,,
\label{bmmap}
\eea
where only the leading $\tan\beta$ dependence is presented, and 
\bea
Br(B_s^0\to\tau\mu)\simeq 3.6\times 10^{-7}\left[\frac{\tan\beta}{60}\right]^8
\left[\frac{100\,\mrm{GeV}}{M_A}\right]^4 \,,
\label{bmtap}
\eea
to be compared with the upper limits in Table~\ref{tab:BLFV}.
In the case of $B_d$ mesons, one should just multiply 
(\ref{bmtap}) with $|V_{td}/V_{ts}|^2 \simeq 0.05$. 
As expected, the Higgs-mediated branching ratio for $B_s^0\to\tau\mu$ 
can be larger than the one for $\tau\to3\mu.$

\begin{table}[t]\begin{center}
\begin{tabular}{|c|c|c|}\hline
Channel & Expt. & Bound (90\% CL)  \\ \hline 
$B_s \to \mu^\pm \mu^\mp$ & CDF~\cite{CDFbmumu}
 & $<2.0 \times 10^{-6}$  \\ \hline
$B_s \to e^\pm \mu^\mp$ & CDF~\cite{CDFbmumu}
 & $<6.1 \times 10^{-6}$  \\ \hline
$B_s \to e^\pm \tau^\mp$ & ---
 & ---   \\ \hline
$B_s \to \mu^\pm \tau^\mp$ & ---
 & ---   \\ \hline
$B_d \to \mu^\pm \mu^\mp$ & BaBar~\cite{BaBar} 
& $<2.0 \times 10^{-7}$  \\ \hline
$B_d \to e^\pm \mu^\mp$ & BaBar~\cite{BaBar} 
& $<2.1 \times 10^{-7}$  \\ \hline
$B_d \to e^\pm \tau^\mp$ & CLEO~\cite{CLEO} 
& $<5.3 \times 10^{-4}$  \\ \hline
$B_d \to \mu^\pm \tau^\mp$ & CLEO~\cite{CLEO}
 & $<8.3 \times 10^{-4}$  \\ \hline
\end{tabular}\end{center}\caption{\it Current experimental 
bounds for the branching ratios of the leptonic $B$ decays  
$B_{s,d}\to l^+_i l^{-}_j$. }
\label{tab:BLFV}
\end{table}

We comment in passing that the decays $B_{d,s}\to \mu e$ are suppressed
by the ratio $m_\mu^2/m_\tau^2 \simeq 0.0036$
compared to $B_{d,s}\to \mu \tau$ and, moreover,  
they are strongly constrained by the process $\mu \to e \gamma$. 
We find that their branching ratios are below the range of 
prospective experimental interest.


{\it 2. Effective Lagrangians and branching ratios for LFV processes.} 
We now present the calculational details we use in 
arriving at the approximate results of the previous Section 
and the numerical results to be presented in the next Section. 
We consider the $R$-parity conserving superpotential:
\begin{eqnarray}
\label{w}
W &=& U^{c}_i (Y_u)_{ij} Q_j H_2
  -  D^{c}_i (Y_d)_{ij}  Q_j H_1 + \nn
&&  N^{c}_i (Y_\nu)_{ij} L_j H_2
  -  E^{c}_i (Y_e)_{ij}  L_j H_1 
  + \frac{1}{2}{N^c}_i (M_N)_{ij} N^c_j + \mu H_2 H_1 \,,
\label{suppot}
\end{eqnarray}
where the indices $i,j$ run over three generations and $M_N$
is the heavy singlet-neutrino mass matrix. We work in a basis where 
$(Y_d)_{ij},$ $(Y_e)_{ij}$ and $(M_N)_{ij}$ are real and diagonal.
At the one-loop level, this leads to the effective 
Lagrangian~\cite{Babu,Isidori,Buras,Babu2}.
\bea
-{\cal L}^{eff}&=&\bar d^i_R Y_{di} \left[ 
\delta_{ij} H_1^0 + \left(\epsilon_0 \delta_{ij} + 
\epsilon_Y \right(Y_u^\dagger Y_u\left)_{ij} \right) H_2^{0\ast }
\right] d^j_L +  h.c. + \nn
&&\bar l^i_R Y_{ei} \left[ 
\delta_{ij} H_1^0 + \left(\epsilon_1 \delta_{ij} + 
\epsilon_2 E_{ij} \right) H_2^{0\ast }
\right] l^j_L + h.c.  \,, 
\label{Leff}
\eea
where $\epsilon_0,$ $\epsilon_Y,$ $\epsilon_1,$ and $\epsilon_2$
are loop-induced form factors, and $E_{ij}=(\Delta m^2_{\tilde 
L})_{ij}/m^2_0$ is the unique source of LFV.
In the following, we present  analytical expressions in the 
mass-insertion approximation, which is simple and known
to reproduce well the full results in the case of large $\tan\beta$. 
However, in  our numerical analyses of the next Section
we use the full diagrammatic  calculation of the $(\bar b s)$-Higgs 
transition of~\cite{Bobeth}, which we modify according to~\cite{Isidori} 
to include the resummed large-$\tan\beta$
contributions to the down-type Yukawa couplings. 
Also, the LFV matrix $E_{ij}$ is calculated exactly using the numerical 
solutions to the full set of renormalization-group
equations of the MSSM, including singlet neutrinos~\cite{h1}.

In the mass-insertion approximation, the down-quark form factors
are~\cite{Isidori,Buras}
\bea
\epsilon_0=\frac{2\alpha_s}{3\pi} \frac{\mu M_{\tilde g}}{m^2_{\tilde d_L}}
F_2\left( x_{\tilde g \tilde d_L},x_{\tilde d_R\tilde d_L} \right) \,, 
\quad\quad
\epsilon_Y=\frac{1}{16\pi^2} \frac{\mu A_{u}}{m^2_{\tilde u_L}}
F_2\left( x_{\mu \tilde u_L},x_{\tilde u_R\tilde u_L} \right) \,,
\eea
where $x_{ab}=m^2_a/m^2_b$ and 
\bea
F_2\left(x,y\right)=
-\frac{x\ln x}{(1-x)(x-y)}-\frac{y\ln y}{(1-y)(y-x)}\,.
\eea
In the lepton sector, the corresponding form factors are~\cite{Babu2}
\bea
\epsilon_{1} &=&
\frac{\alpha'}{8\pi}\mu M_1 \left[2F_3\left(M_1^2,m_{\tilde l_L}^2,
m_{\tilde l_R}^2\right) -
F_3\left(M_1^2,\mu^2,m^2_{\tilde l_L}\right) + 
2F_3\left(M_1^2,\mu^2,
m^2_{\tilde l_R}\right)\right] + \nonumber \\ 
&&
\frac{\alpha_2}{8\pi}\mu M_2\left[F_3\left(\mu^2,
m^2_{\tilde l_L}, M_2^2\right) 
+ 2F_3\left(\mu^2,m_{\tilde \nu}^2,M_2^2\right)\right]\,,
\label{eps1} 
\\
\epsilon_{2} &=&  
\frac{\alpha'}{8\pi} m_0^2 \mu M_1 
\left[
2F_4\left(M_1^2,m_{\tilde l_L}^2,
m_{\tilde \tau_L}^2,m_{\tilde \tau_R}^2\right)
-F_4 \left(\mu^2, m^2_{\tilde l_L}, m^2_{\tilde \tau_L},M_1^2\right)
\right] + \nn
&&
\frac{\alpha_2}{8\pi}m_0^2 \mu M_2 
\left[
F_4\left(\mu^2,m^2_{\tilde l_L}, m^2_{\tilde \tau_L}, M_2^2\right) +
2F_4\left(\mu^2,m_{\tilde\nu_l}^2,m_{\tilde \nu_\tau}^2,M_2^2\right)
\right] \,,
\eea
where
\bea
F_3\left(x,y,z\right)&=& 
-\frac{xy\ln (x/y)+yz\ln (y/z)+zx\ln (z/x)}
{(x-y)(y-z)(z-x)} \,,  \nn
F_4\left(x,y,z,w\right) &=&
-\frac{x \ln x}{(x-y)(x-z)(x-w)} -\frac{y\ln y}{(y-x)(y-z)(y-w)}+ \\
&&(x\leftrightarrow z,
y\leftrightarrow w)\, . \nonumber
\eea
The resulting effective Lagrangians describing flavour-violating neutral
Higgs interactions with down quarks and charged leptons 
are~\cite{Isidori,Babu2}
\bea
-{\cal L}^{eff}_{d_i\neq d_j} =
(2G_F^2)^{1/4}
\frac{m_{d_i} \kappa^d_{ij}}{\cos^2\beta}
\left(\bar d_{iR}\,d_{jL}\right)
\left[\cos(\alpha-\beta) h^0 + \sin(\alpha-\beta) H^0 - iA^0 \right]+h.c.\,,&& 
\label{Leffd} \\
-{\cal L}^{eff}_{l_i\neq l_j} =
(2G_F^2)^{1/4}
\frac{m_{l_i} \kappa^l_{ij}}{\cos^2\beta}
\left(\bar l_{iR}\,l_{jL}\right)
\left[\cos(\alpha-\beta) h^0 + \sin(\alpha-\beta) H^0 - i A^0\right]+h.c.
\,,\,\,\,&&
\label{Leffl}
\eea
where 
\bea
\kappa^d_{ij} &=& \frac{\eps_Y Y^2_{t} \bar\lambda^t_{ij}}
{\left[1+(\eps_0+\eps_Y Y^2_{t} \delta_{it})\tan\beta\right]
\left[1+\eps_0 \tan\beta\right] }\,, \\
\kappa^l_{ij} &=& \frac{\eps_2 E_{ij}}{
\left[1+(\eps_1+\eps_2
E_{ii})\tan\beta\right]^2 }\,.
\eea
As we are interested in $B$-meson decays, we have
$\bar\lambda^t_{bq}=V^\ast_{tb} V_{tq},$ where the $V_{ij}$ are CKM 
matrix elements. 

Using \rfn{Leffl}, one easily obtains the branching ratio for 
the decay $\tau\to\mu\mu\mu$, given by
\bea
Br(\tau\to3\mu) &=&
\frac{G_F^2 m_\mu^2 m_\tau^7 \tau_\tau}{1536\pi^3 \cos^6\beta}
 |\kappa_{\tau\mu}^l|^2 
\left[\left(\frac{\sin(\alpha-\beta)\cos\alpha}{M_{H^0}^2} -
\frac{\cos(\alpha-\beta)\sin\alpha}{M_{h^0}^2}\right)^2 
+\frac{\sin^2\beta}{M_A^4}\right]
\nn
&\approx& 
\frac{G_F^2 m_\mu^2 m_\tau^7 \tau_\tau}{768\pi^3 M_A^4}
 |\kappa_{\tau\mu}^l|^2 \tan^6\beta \,,
\eea
where $\tau_\tau$ is the $\tau$ lifetime and the large-$\tan\beta$
limit is taken in the last step. This is the result that was used to 
derive the estimate \rfn{t3map}.

The branching ratio for $\tau\to\mu\gamma$ in the mass-insertion 
approximation reads~\cite{h1}:
\bea
Br(\tau\to \mu\gamma) = \frac{\alpha}{4}m^5_\tau \tau_\tau |A_L|^2 \,,
\eea
where
\bea
A_L & \approx & (\Delta m^2_{\tilde L})_{\tau\mu}
\frac{\alpha_2}{4\pi} \mu M_2\tan\beta \nn
& \times &
D\left[
D\left[\frac{1}{m^2}
\left\{ f_c(x_{Mm})- \frac{1}{4} f_n(x_{Mm}) 
\right\}; M^2 \right] 
(M_2^2,\mu^2);m^2\right]
(m^2_{\tilde \nu_\mu}, m^2_{\tilde \nu_\tau})\,,
\eea
where
$D[f(x);x](x_1,x_2)= (f(x_1) - f(x_2))/(x_1-x_2) ,$
and
\bea
f_c(x) &=& -\frac{1}{2(1-x)^3} (3-4 x + x^2 + 2\ln x)\,, \nn
f_n(x) &=& \frac{1}{(1-x)^3} (1- x^2 + 2x\ln x)\,. \nonumber
\eea
This is the result that was used to derive the estimate \rfn{tmgap}.

The dominant operators at large $\tan\beta$ 
in the effective Hamiltonian describing 
the transition $\bar b\to \bar s l_i^+ l_j^-$ are
\br
{\cal H}=-\frac{G_F^2 M^2_W}{\pi^2} V_{tb}^*V_{ts} 
\biggl [ c_S^{ij} {\cal O}_S^{ij}
+c_P^{ij} {\cal O}_P^{ij} + c_{10}^{ij} {\cal O}_{10}^{ij} \biggr ] 
+ h.c. \;, 
\label{ham}
\er
where $V$ is the CKM matrix and $G_F$ the Fermi coupling constant.
Here $c_{S,P,10}$ and ${\cal  O}_{S,P,10}$ are, respectively, the Wilson 
coefficients and the local operators~\footnote{In (\ref{ham}) we have
omitted operators proportional to the strange-quark mass, since they
are subleading for our processes.}, which are given by
\br
{\cal O}_{10}^{ij} &=& 
(\bar{b}_R\gamma^\mu s_L)(\bar{l}_i\gamma_\mu \gamma_5 l_j) \nonumber \;,  \\  
{\cal O}_S^{ij} &=& m_b (\bar{b}_R s_L) \bar{l}_i l_j 
\nonumber \;, \\
 {\cal O}_P^{ij} &=&  m_b (\bar{b}_R s_L) \bar{l}_i \gamma_5 l_j 
\;.
\er
In the SM with the seesaw mechanism, the only non-zero  
coefficients are the $c_{10}.$ For $i\neq j$, they scale
like the square of the inverse mass of the singlet neutrinos,
and are completely negligible. In supersymmetric 
seesaw models, there are two additional operators in (\ref{ham}), 
the scalar ${\cal O}_S$ and the pseudoscalar ${\cal O}_P$,
whose coefficients dominate
over the SM contributions. In particular, for $i\neq j$ they
are suppressed by the scale of supersymmetry-breaking soft masses only.
At large $\tan\beta$ the dominant contributions to $c_S^{ij}$ and 
$c_P^{ij}$ come from the penguin and double-penguin
diagrams  presented in Fig.~\ref{fig1} and Fig.~\ref{fig2},
respectively.
We should note here, however, that when the Higgs masses are ${\cal 
O}$(TeV), 
and the sneutrinos are very light, the box diagrams may dominate. 
In this case we obtain in general small branching ratios,
$Br(B_s \to \mu \tau) \lsim 10^{-14}$, when we take 
the $\tau\to \mu \gamma$ constraint into account.

By setting $c_{10}$ to zero in (\ref{ham}), we obtain the 
following branching
ratio~\footnote{We note that (\ref{brll}) can be straightforwardly 
extended to $B_d$ LFV decays by  replacing $s \to d$.}, 
\br
{\rm Br}(B_s\to l_i l_j) &=& \frac{G_F^4 M^4_W}{8\pi^5}
|V_{tb}^*V_{ts}|^2 M_{B_s}^5 f_{B_s}^2 \tau_{B_s} \biggl (\frac{m_b}{m_b+
m_s}\biggr )^2 \nonumber \\[3mm] &\times & 
\sqrt{ \biggl [1-\frac{(m_{l_i}+m_{l_j})^2}{M_{B_s}^2}\biggr ]
\biggl [1-\frac{(m_{l_i}-m_{l_j})^2}{M_{B_s}^2}\biggr ] }       
\nonumber \\[3mm] &\times & 
\Biggl \{ \biggl (1-\frac{(m_{l_i}+m_{l_j})^2}{M_{B_s}^2}\biggr ) 
|c^{ij}_{S}|^2
+\biggl ( 1-\frac{(m_{l_i}-m_{l_j})^2}{M_{B_s}^2}\biggr ) |c_{P}^{ij}|^2
\Biggr \} \;,
\label{brll}
\er
where $M_{B_s}$ and $\tau_{B_s}$ are  the mass and lifetime of the 
$B_s$ meson, and $f_{B_s}=230\pm 30$ GeV~\cite{fbs} is its decay constant.

For $B_s\to \mu\mu$ decay, the form factors are~\cite{Isidori,Buras}:
\begin{eqnarray}
c_S^{\mu\mu} &=&
\frac{\sqrt{2}\pi^2}{G_F M^2_W}
\frac{m_\mu\kappa_{bs}^d}{\cos^3\beta \bar\lambda^t_{bs}}
\left[{\sin(\alpha-\beta)\cos\alpha\over M^2_{H^0}}-
{\cos(\alpha-\beta)\sin\alpha\over M^2_{h^0}}  \right]  \nn
&\approx&
-\frac{4\pi^2 m_\mu  m_t^2}{ M^2_W}
\frac{ \epsilon_Y ~\tan^3\beta }
{\left[1+(\eps_0+\eps_Y Y^2_{t})\tan\beta\right]
\left[1+\eps_0 \tan\beta\right]}
\left[{1\over M^2_{A^0}}\right] \,,
\label{cs} \\[3mm]
c_P^{\mu\mu} &=&
-\frac{\sqrt{2}\pi^2}{G_F M^2_W}
\frac{m_\mu\kappa_{bs}^d}{\cos^3\beta \bar\lambda^t_{bs}}
\left[{\sin\beta\over M^2_{A^0}} \right] 
\approx c_S^{\mu\mu} \,,
\label{cp}
\end{eqnarray}
and, for the double-penguin contribution to $B_s\to \mu\tau$ decay, we 
obtain 
from \rfn{Leffd} and \rfn{Leffl} the form factors
\bea
c_S^{\mu\tau}=c_P^{\mu\tau} &=&
\frac{\sqrt{2}\pi^2}{G_F M^2_W}
\frac{m_\tau\kappa_{bs}^d \kappa_{\tau\mu}^{\ast l}}
{\cos^4\beta \bar\lambda^t_{bs}}
\left[{\sin^2(\alpha-\beta)\over M^2_{H^0}}+
{\cos^2(\alpha-\beta)\over M^2_{h^0}} + {1\over M^2_{A^0}} \right]
\nonumber \\[3mm]
&\approx&
\frac{8\pi^2 m_\tau  m_t^2}{ M^2_W}
\frac{ \epsilon_Y~ \kappa_{\tau\mu}^{\ast l} ~\tan^4\beta }
{\left[1+(\eps_0+\eps_Y Y^2_{t} )\tan\beta\right]
\left[1+\eps_0 \tan\beta\right]}
\left[{1\over M^2_{A^0}}\right] \,.
\label{ctm}
\eea
The last two results are those that were used to derive the estimates 
\rfn{bmmap} and \rfn{bmtap}. In the calculation for the ratio
$B_s\to \mu\tau$ one also has contributions from the operators
$(\bar{b}P_{L(R)}s)(\bar{\mu}P_{L(R)}\tau)$. However, their
contribution is proportional
to $\left[{\sin^2(\alpha-\beta)\over M^2_{H^0}}+
{\cos^2(\alpha-\beta)\over M^2_{h^0}} - {1\over M^2_{A^0}} \right]$
which vanishes approximately at large $\tan\beta$. 
Furthermore,
the operator $(\bar{b}P_R s)(\bar{\mu}P_L\tau)$ is proportional
to $m_s m_\mu$ and thus subdominant to $(\bar{b}P_{L}s)(\bar{\mu}P_{R}\tau)$
we consider here. For the same reason, the leptonic CP-asymmetries 
in $B_s\to \mu\tau$ decays due to the complex $\kappa^l_{\tau\mu}$ 
in \rfn{ctm} are of order $m_\mu/m_\tau.$

{\it 3. Numerical Analyses.}
The purpose of this work is to study in a complete way the allowed rates 
for the Higgs-mediated LFV processes in supersymmetric seesaw models 
in which the only source of LFV is the renormalization of the soft 
supersymmetry-breaking mass parameters above $M_{N_i}$, due to the 
singlet-neutrino 
Yukawa couplings. We work with two models. {\it First}, we study the 
constrained
MSSM (CMSSM) which has just two universal mass parameters at GUT scale,
$M_{1/2}$ for gauginos and $m_0$ for all scalars, taking for simplicity
$A_0=0$ and not expecting our results to be sensitive to its exact value.
{\it Secondly}, we study the more
general flavour-universal MSSM (GFU-MSSM) in which the universal masses
for squarks, sleptons and the Higgs doublets $H_1$ and  $H_2$ 
are different from each other.
This permits different mass scales for squarks and sleptons which,
in turn, are independent of the Higgs boson masses. However, we always 
require that squark and slepton mass matrices at the GUT scale are 
each proportional
to unit matrices. If one goes beyond this assumption, 
arbitrary sources of flavour violation appear in the soft 
supersymmetry-breaking sector, and
the model loses all the predictivity, in particular the connection between 
the LFV and the neutrino masses and mixings. Moreover, LFV rates generally 
exceed experimental limits~\cite{ENBG}.

We parametrize the singlet-neutrino Yukawa couplings $Y_\nu$ and masses
$M_{N_i}$ in terms of low energy neutrino data according to~\cite{par}.
We generate all the free parameters of the model randomly and 
calculate the low-energy sparticle masses and mixings by solving 
numerically
the one-loop renormalization-group equations and imposing the requirement 
of
electroweak symmetry breaking. 
We calculate the rates for the LFV processes as described in the last 
Section. For the decays $l_i\to l_j\gamma$, we use 
the exact diagrammatic formulae in~\cite{h1}. 

We fix the light-neutrino parameters by~\cite{Giunti} $\Delta
m^2_{32}=3\times 10^{-3}$ eV$^2,$ $\Delta m^2_{21}=5\times 10^{-5}$
eV$^2,$ $\tan^2\theta_{23}=1$, $\tan^2\theta_{12}=0.4$ and
$\sin\theta_{13}=0.1$, corresponding to the LMA solution for the solar
neutrino anomaly. The neutrino mixing phase is taken to be maximal:
$\delta=\pi/2.$ The overall light-neutrino mass scale is randomly
generated in the range $(10^{-6} \to 10^{-1})$ eV, with a constant
logarithmic measure.  We assume the normal mass hierarchy, which is
favoured by leptogenesis arguments~\cite{lepto}. The rest of the
neutrino parameters are calculated from the randomly generated textures
$H_1$ (which maximizes $\tau-\mu$ violation) and $H_2$ (which maximizes
$\tau-e$ violation), that were introduced in~\cite{par}.

We fix $\tan\beta=60$, which is the largest value for which electroweak
symmetry breaking is obtained generically. In the case of the CMSSM, the
mass parameters $M_{1/2}$ and $m_0$ for sleptons, squarks and both Higgses
are generated randomly in the range $(0 \to 700)$ GeV. In the case of the
GFU-MSSM, we have four scalar mass parameters $m_0^{\tilde q},$
$m_0^{\tilde l},$ $m_0^{H_1}$ and $m_0^{H_2}$ for squarks, sleptons and
Higgs doublets, respectively. We generate each of them randomly and
independently in the range $(0 \to 700)$ GeV. We impose the experimental
upper bounds $Br(\tau\to\mu\gamma)<2\times 10^{-6},$ $Br(\tau\to
e\gamma)<2\times 10^{-6}$ and $Br(\mu\to e\gamma)<1.2\times 10^{-11}$, and
the lower bounds on unobserved particle masses are taken to be $m_{\tilde
l},m_{\tilde q},M_A>100$ GeV, and no other phenomenological bounds are
imposed. Thus, our analysis should give an accurate comparison of
different LFV decays and $B_{s,d}\to\mu\mu$ with each other, but may be
somewhat optimistic concerning the overall rates.

\begin{figure}[htbp]
\centerline{\epsfxsize = 0.5\textwidth \epsffile{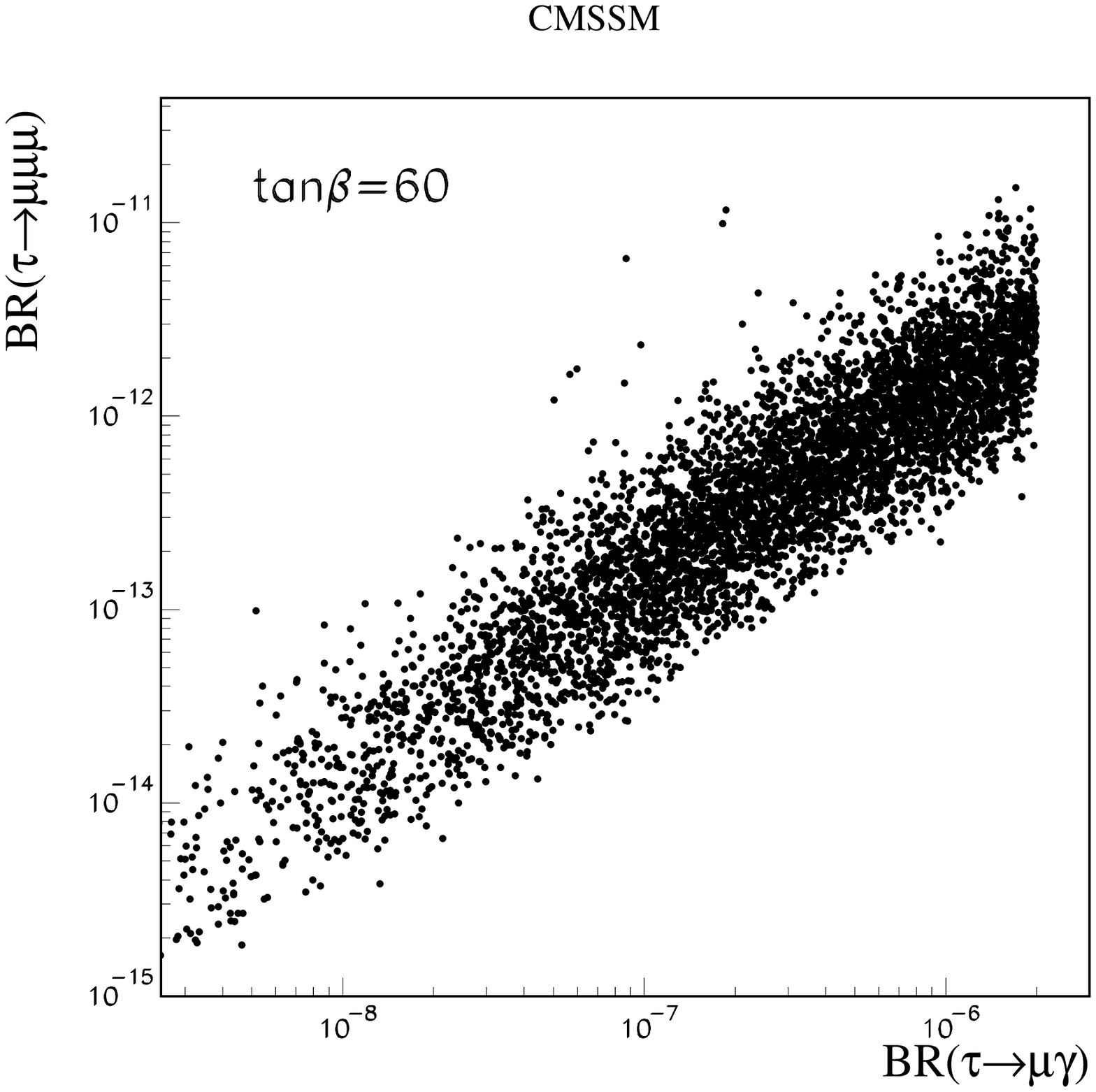} 
\hfill \epsfxsize = 0.5\textwidth \epsffile{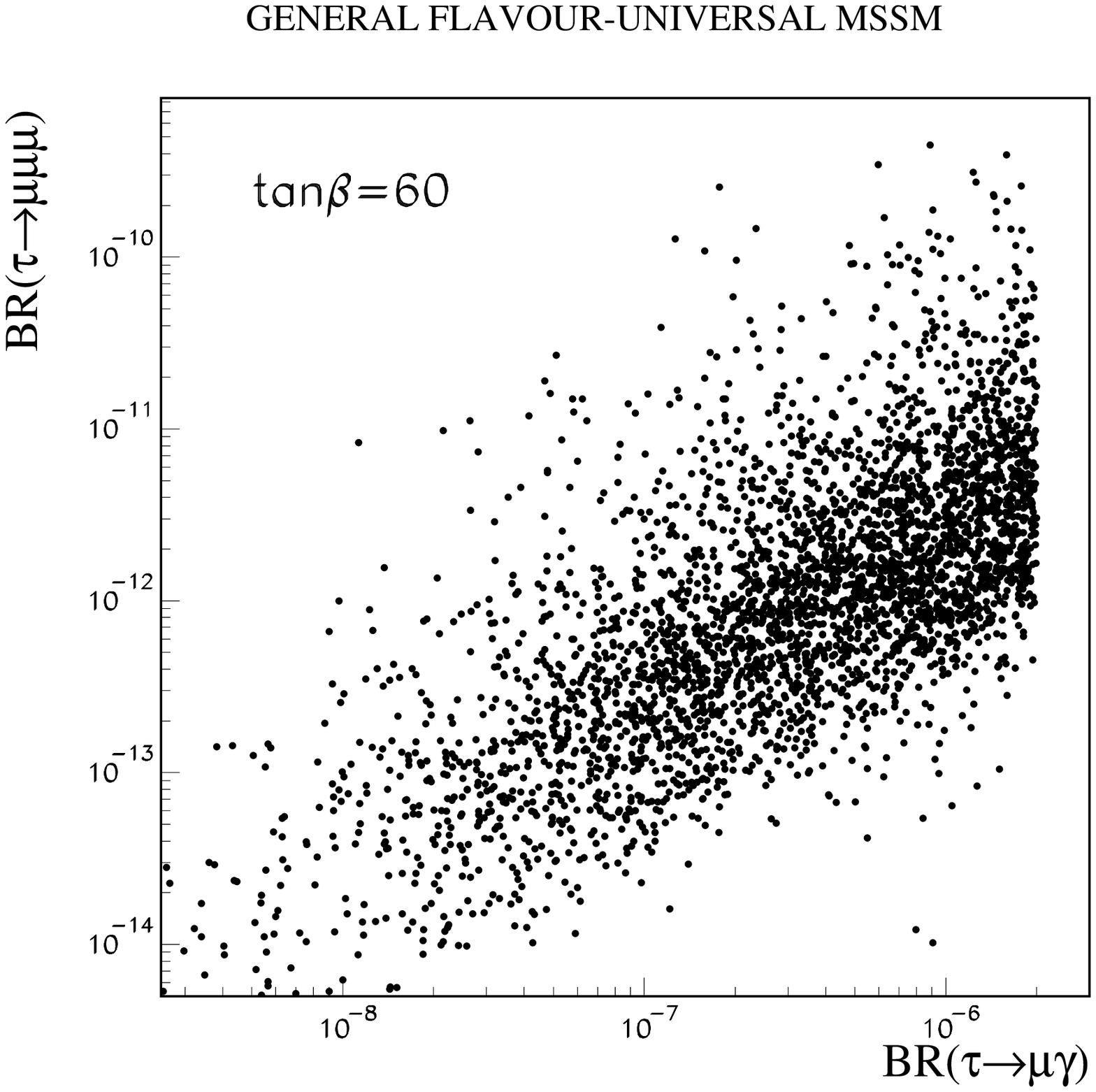} 
}
\caption{\it 
Scatter plot of Higgs-mediated $Br(\tau\to 3\mu)$ against $Br(\tau\to 
\mu\gamma)$ in
(a) the CMSSM and (b) the GFU-MSSM. 
\vspace*{0.5cm}}
\label{fig3}
\end{figure}

\begin{figure}[htbp]
\centerline{\epsfxsize = 0.5\textwidth \epsffile{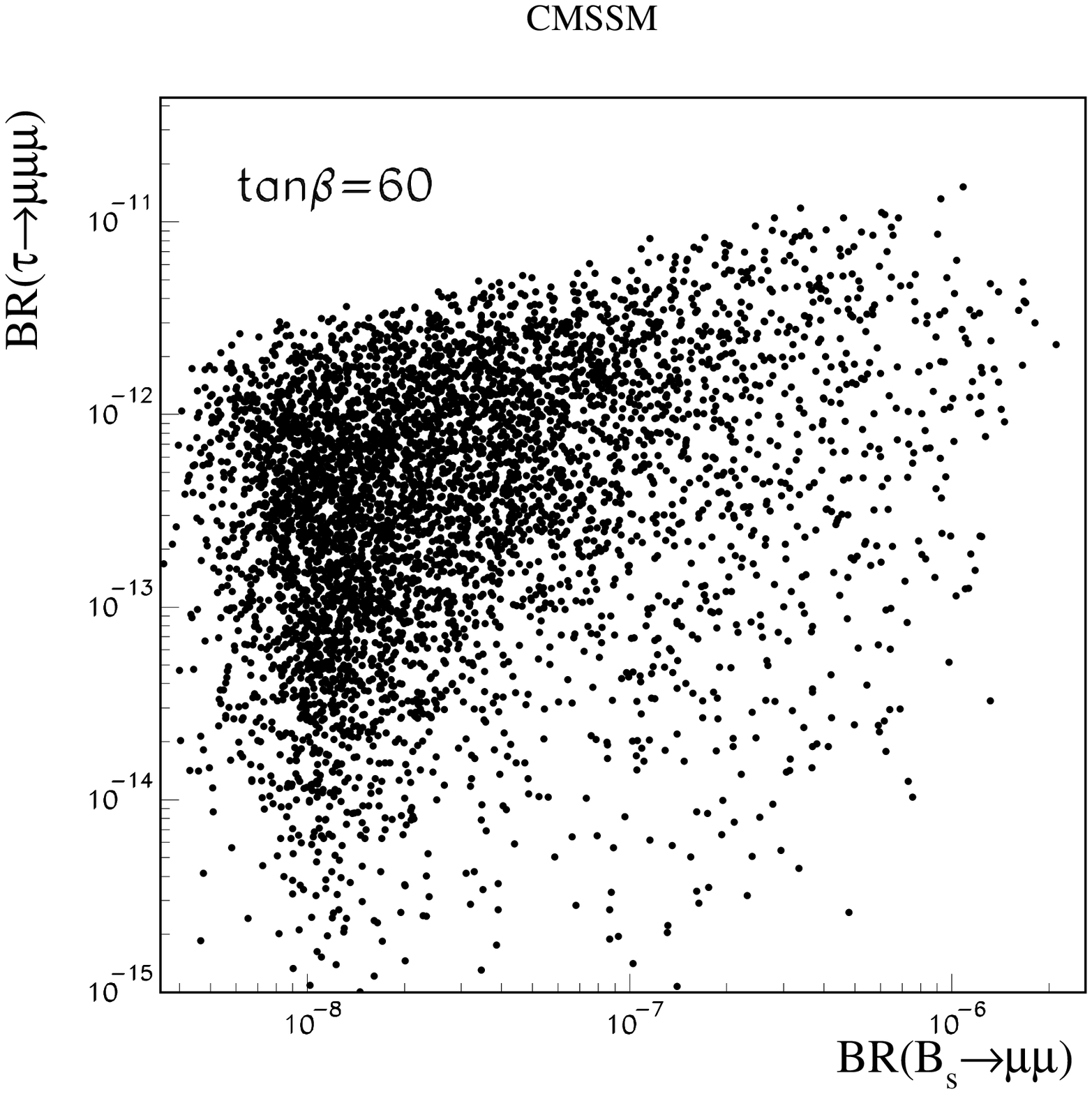} 
\hfill \epsfxsize = 0.5\textwidth \epsffile{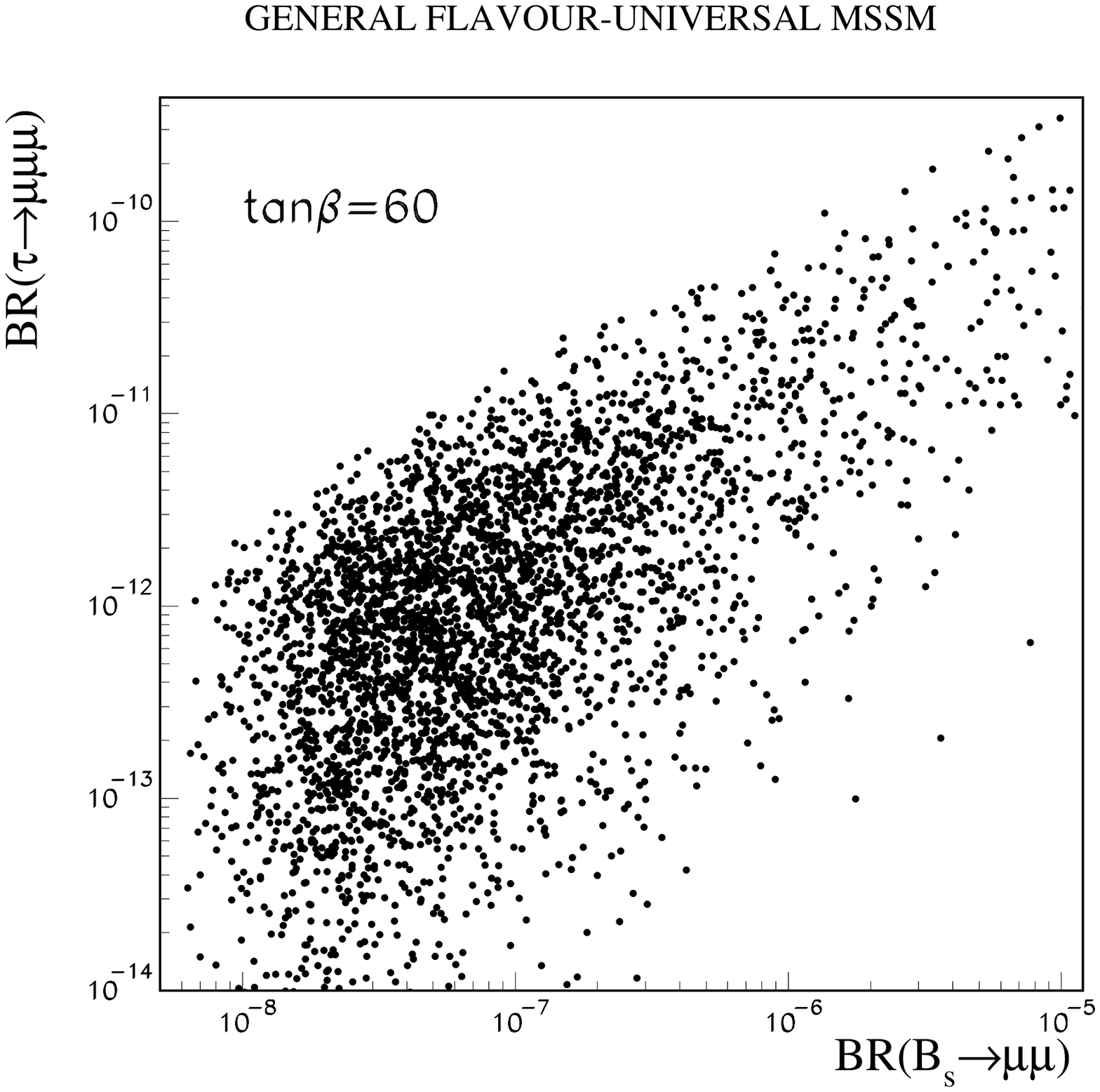} 
}
\caption{\it 
Scatter plot of Higgs-mediated $Br(\tau\to 3\mu)$ against $Br(B_s\to 
\mu\mu)$ in
(a) the CMSSM and (b) the GFU-MSSM. 
\vspace*{0.5cm}}
\label{fig4}
\end{figure}

In Fig.~\ref{fig3} we present scatter plots of Higgs-mediated $Br(\tau\to
3\mu)$ against $Br(\tau\to \mu\gamma)$ in (a) the CMSSM and (b) the
GFU-MSSM. Whilst $Br(\tau\to 3\mu)<10^{-11}$ in the CMSSM, in the GFU-MSSM
the present $Br(\tau\to \mu\gamma)$ bound allows $Br(\tau\to 3\mu)<4\times
10^{-10}.$ This increase could have been expected, since in the GFU-MSSM
the Higgs masses and slepton masses are independent of each other. The
points with the highest possible $Br(\tau\to 3\mu)$ in In Fig.~\ref{fig3}
correspond to $M_A\approx 100$ GeV. However, for $\tan\beta=60$, such low
values for $M_A$ are constrained by $B_s\to\mu\mu.$ To see this, we plot
in Fig.~\ref{fig4} $Br(\tau\to 3\mu)$ against $Br(B_s\to\mu\mu)$ in both
models we consider. It follows that, in the GFU-MSSM, the present bound
$Br(B_s\to\mu\mu)<2\times 10^{-6}$ constrains $Br(\tau\to 3\mu)<1\times
10^{-10}.$ This should be compared with the rate for the photonic penguin
contribution to $Br(\tau\to 3\mu)$, which is given by \rfn{ph}. Our 
results
show that, {\it after imposing the $\tau\to \mu\gamma$ and 
$B_s\to \mu\mu$ bounds}, the 
photonic penguin dominates over the Higgs-mediated contribution.

We emphasize that we have been seeking to maximize LFV effects in these
figures. We also note that the lower bounds on the LFV processes in the
plots are artificial, being due to the lower bounds on the generated
values of $Y_\nu$.

\begin{figure}[htbp]
\centerline{\epsfxsize = 0.5\textwidth \epsffile{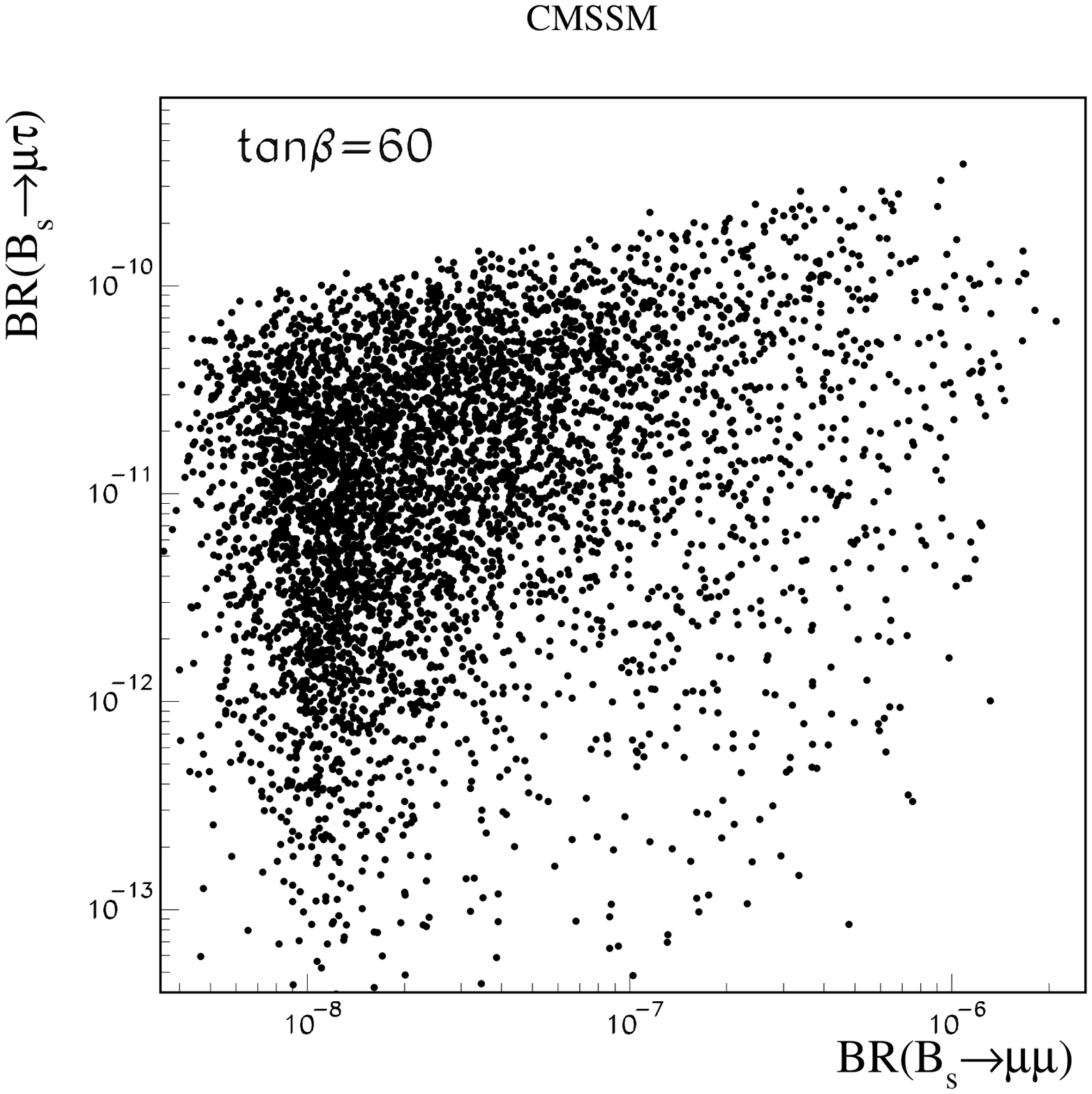} 
\hfill \epsfxsize = 0.5\textwidth \epsffile{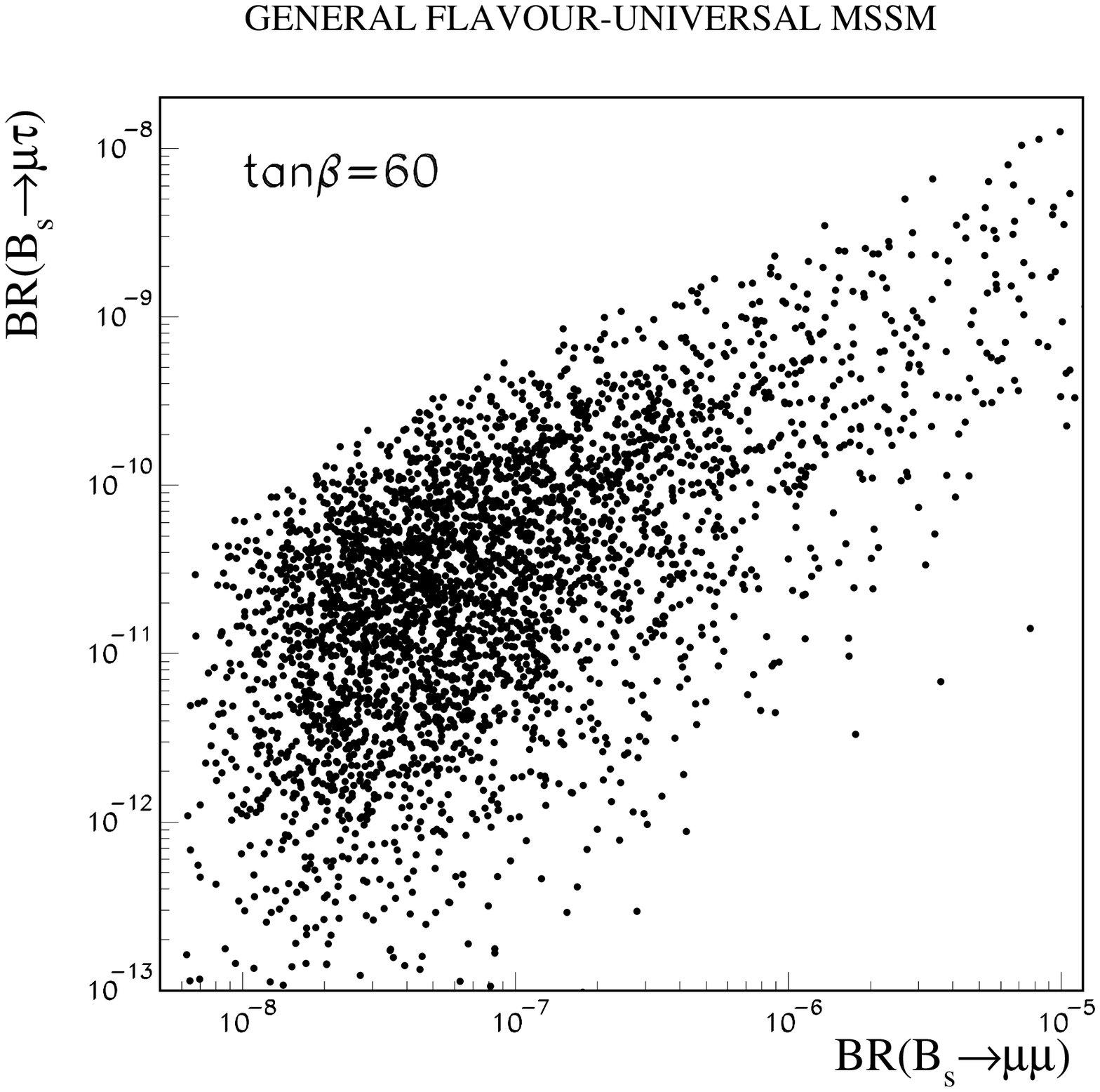} 
}
\caption{\it 
Scatter plot of Higgs-mediated $Br(B_s\to \mu\tau)$ against $Br(B_s\to 
\mu\mu)$ in
(a) the CMSSM and (b) the GFU-MSSM. 
\vspace*{0.5cm}}
\label{fig5}
\end{figure}

\begin{figure}[htbp]
\centerline{\epsfxsize = 0.5\textwidth \epsffile{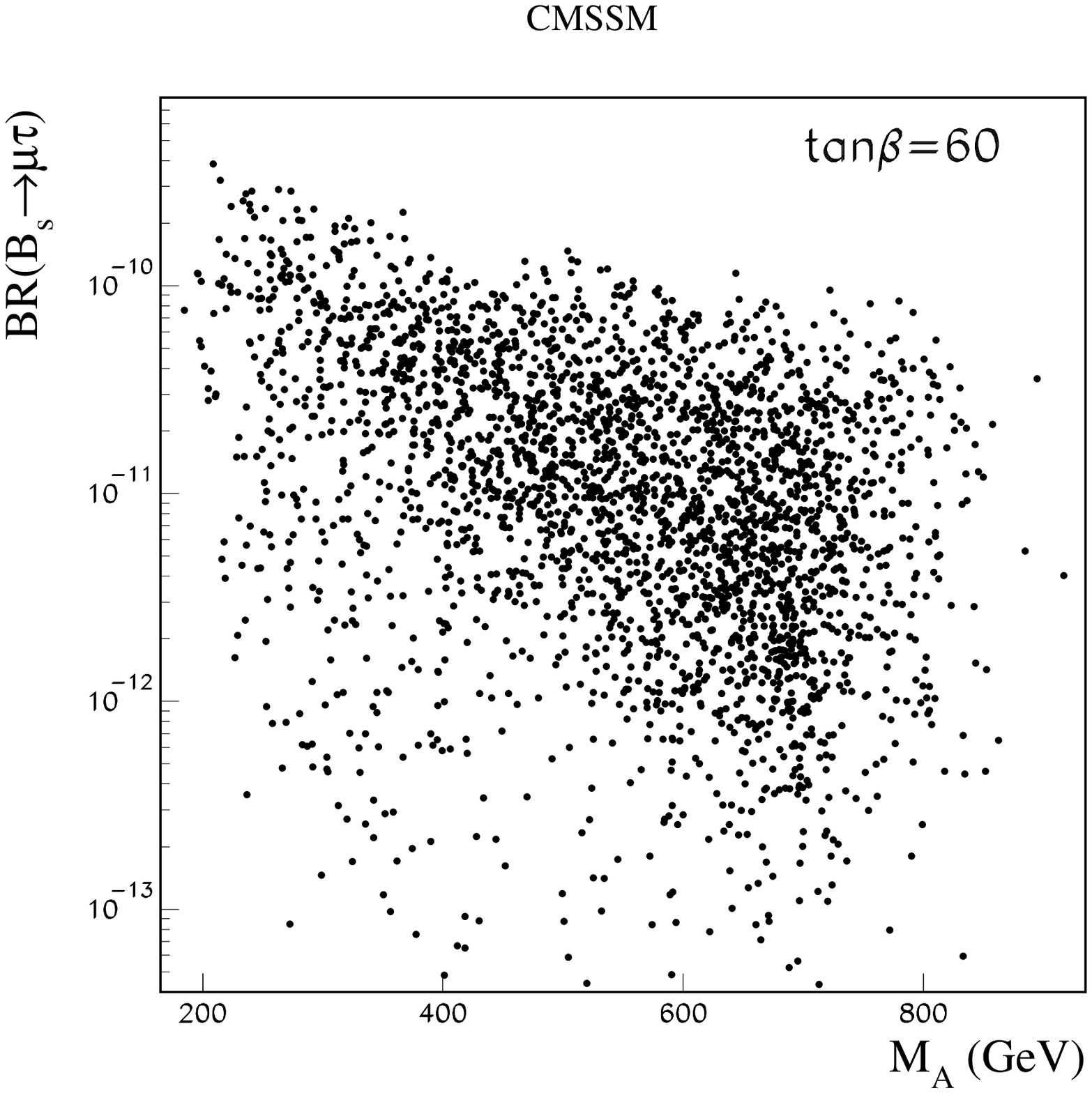} 
\hfill \epsfxsize = 0.5\textwidth \epsffile{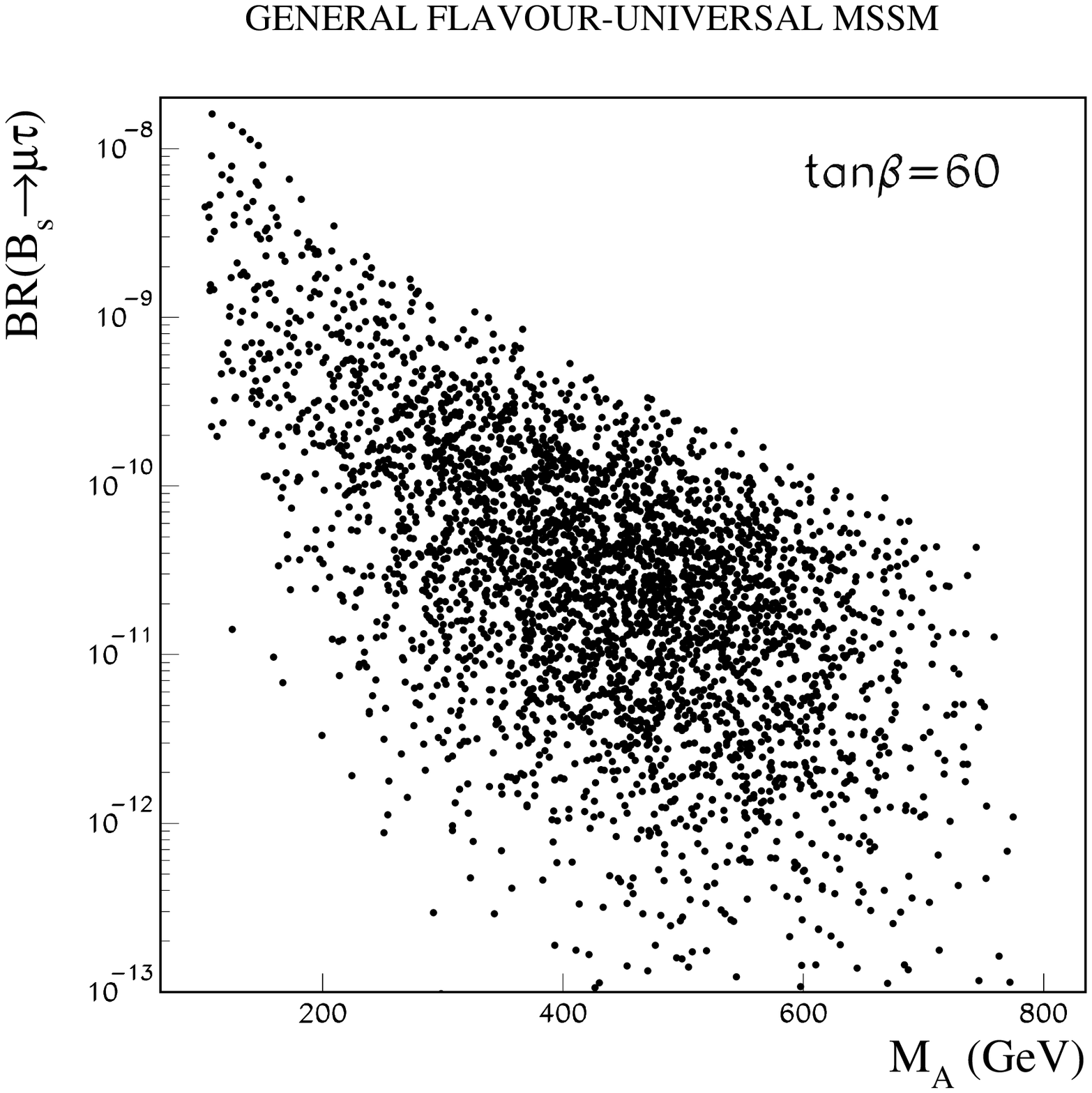} 
}
\caption{\it 
Scatter plot of Higgs-mediated $Br(B_s\to \mu\tau)$ against the 
pseudoscalar 
Higgs mass $M_A$ in (a) the CMSSM and (b) the GFU-MSSM. 
\vspace*{0.5cm}}
\label{fig6}
\end{figure}

Proceeding with the $B$-meson decays, we plot in Fig.~\ref{fig5} the
Higgs-mediated $Br(B_s\to \mu\tau)$ against $Br(B_s\to \mu\mu)$ in (a) the
CMSSM and (b) the GFU-MSSM. Again, there is a sharp upper bound on
$Br(B_s\to \mu\tau)$ in both models. In the GFU-MSSM the maximum allowed
value for $Br(B_s\to \mu\tau)$ is $4\times 10^{-9}$, which is larger than
that for $\tau\to 3\mu.$ However, detecting $\tau$ leptons is
experimentally challenging, particularly in a high-rate environment such
as the LHC. To see the dependence of $B_s\to \mu\tau$ on $M_A$, we plot in
Fig.~\ref{fig6} the value of $Br(B_s\to \mu\tau)$ against $M_A.$ We
emphasize that $M_A$ can be very small in the GFU-MSSM~\cite{EFOS}, whilst
in the CMSSM there is an approximate lower bound $M_A>200$ GeV, because it
is related to the slepton masses. The points with the largest $Br(B_s\to
\mu\tau)$ in Fig.~\ref{fig6}, as well as the points with the largest
$Br(\tau\to 3\mu)$ in Fig.~\ref{fig3},~\ref{fig4}, correspond to
$M_A={\cal O}(100)$ GeV.

Similar results are also valid for the decays $\tau\to e\mu\mu$ and
$B_s\to e\tau$. Since phenomenologically the $\tau-e$ transition can be as
large as the $\tau-\mu$ one~\cite{par}, we have the same bounds
$Br(\tau\to e\mu\mu) < 4\times 10^{-10}$ and $Br(B_s\to e\tau) < 4\times
10^{-9}.$ The scatter plots for those processes are indistinguishable from
Figs.~\ref{fig3} to \ref{fig6}, so we do not present them here. We have
also studied the decays $B_{s,d}^0\to\mu e$ and $\mu\to eee.$ The
Higgs-mediated contributions to $Br(B_{s,d}^0\to\mu e)$ and $Br(\mu\to
eee)$ are suppressed by the $\mu\to e\gamma$ constraint and small Yukawa
couplings to be below $10^{-15}$ and $10^{-21},$ respectively.


{\it 4. Conclusions.}

In this Letter we have studied the allowed rates for Higgs-mediated LFV
decays in supersymmetric seesaw models, assuming that LFV is generated
entirely by the renormalization effects of neutrino Yukawa couplings. Even
if we allow the mass scales for squarks, sleptons and two Higgs doublets
to differ from each other, the bounds due to the decays
$\tau\to\mu\gamma,$ $\tau\to e\gamma,$ $\mu\to e\gamma$ and $B_s\to
\mu\mu$ are very constraining. We obtain the following bounds on the
Higgs-mediates processes: $Br(B_{s}^0\to\mu\tau, e\tau)\lsim 4\times
10^{-9}$ and $Br(\tau\to \mu\mu\mu, e\mu\mu)\lsim 4\times 10^{-10}$. We
have discussed these numerical results in terms of approximate analytical
expressions, but our calculations are more exact. We conclude that the
Higgs-mediated contributions to $\tau\to \mu\mu\mu, e\mu\mu$ are
subleading compared to the photonic penguin ones.

\vskip 0.5in
\vbox{
\noindent{ {\bf Acknowledgements} } \\
\noindent  
A.D would like to thank Maxim Pospelov for illuminating discussions
during the SUSY~02 conference in Hamburg. 
This work was partially supported by EU TMR
contract No.  HPMF-CT-2000-00460, by ESF grant No. 5135. 
A.D. would also like to acknowledge financial support from the CERN Theory 
Division and
the Network RTN European Program HPRN-CT-2000-00148 `Physics Across
the Present Energy Frontier: Probing the Origin of Mass'. 
}

\newpage


\end{document}